\documentclass[acmsmall,screen]{acmart}

\usepackage[switch]{lineno}
\usepackage{cuted}
\newcommand{\toolname}{\textsc{ReProAgent}\xspace}
\usepackage{mdframed}

\usepackage{bbding}

\usepackage{tcolorbox}
\tcbuselibrary{breakable}
\usepackage{pifont}
\usepackage{enumerate}
\usepackage{ragged2e}
\usepackage{makecell}
\usepackage{stfloats}
\usepackage[vlined, ruled]{algorithm2e}
\usepackage{algorithmic}
\usepackage{graphicx}
\usepackage{textcomp}
\usepackage{multirow}
\usepackage{subfigure}
\usepackage{diagbox}
\usepackage{listings}
\usepackage{float}
\usepackage{url}
\usepackage{graphicx}

\usepackage{hyperref}
\definecolor{deepblue}{rgb}{0,0,0.5}
\definecolor{deepgreen}{rgb}{0,0.5,0}
\definecolor{deepred}{rgb}{0.6,0,0}
\definecolor{darkorange}{RGB}{255,140,0}
\definecolor{lightgray}{rgb}{0.93,0.93,0.93}
\definecolor{deepgray}{rgb}{0.25,0.25,0.25}
\hypersetup{
    colorlinks=true,
    linkcolor=blue,
    filecolor=blue,      
    urlcolor=blue,
    citecolor=blue,
}

\usepackage{multirow}
\usepackage[utf8]{inputenc}
\usepackage{graphicx}
\usepackage{textcomp}
\usepackage{color,xcolor}
\usepackage{url}
\usepackage{graphicx}
\usepackage{subfigure}
\usepackage{stmaryrd}
\usepackage{verbatimbox}
\usepackage{enumerate}
\usepackage[shortlabels]{enumitem}
\usepackage{bm}

\usepackage{makecell}
\usepackage{booktabs}

\usepackage{multirow}

\usepackage{enumitem}
\usepackage{colortbl}

\newcommand{\finding}[2]{
\begin{center}
\begin{tcolorbox}[leftrule=0mm,toprule=0mm,bottomrule=0mm,rightrule=0mm,left=1pt,right=2pt,top=0pt,bottom=0pt,breakable]
\textbf{Answer to RQ{#1}:}
{#2}
\end{tcolorbox}
\end{center}
}

\newcommand{\summary}[1]{
\begin{center}
\begin{tcolorbox}[leftrule=0mm,toprule=0mm,bottomrule=0mm,rightrule=0mm,left=1pt,right=2pt,top=0pt,bottom=0pt,breakable]
\textbf{Summary:}
{#1}
\end{tcolorbox}
\end{center}
}

\usepackage[normalem]{ulem}

\newcommand{\delete}[1]{}

\AtBeginDocument{
  \providecommand\BibTeX{{
    \normalfont B\kern-0.5em{\scshape i\kern-0.25em b}\kern-0.8em\TeX}}}

\setcopyright{acmcopyright}
\copyrightyear{2026}
\acmYear{2026}
\acmDOI{1}

\acmVolume{1}
\acmNumber{1}
\acmArticle{1}
\acmMonth{1}

\begin{document}

\title{\toolname{}: Tool-Augmented Multi-Stage Agentic Generation of Bug Reproduction Tests from Issue Reports}

\author{Quanjun Zhang}
\email{quanjunzhang@njust.edu.cn}
\orcid{0000-0002-2495-3805}
\affiliation{
\institution{School of Computer Science and Engineering, Nanjing University of Science and Technology}
\city{Nanjing}\country{China}
}

\author{Yi Zheng} 
\orcid{0009-0009-2040-8496}
\email{201250182@smail.nju.edu.cn}

\author{Ye Shang}
\email{yeshang@smail.nju.edu.cn}
\orcid{0009-0000-8699-8075}

\affiliation{
\institution{State Key Laboratory for Novel Software Technology, Nanjing University}
\institution{Nanjing University}
\city{Nanjing}\country{China}
}

\author{Weifeng Sun}
\email{wfsun@smu.edu.sg}
\orcid{0000-0001-6013-1369}
\affiliation{%
	\institution{Singapore Management University}
	\country{Singapore}
}

\author{Haichuan Hu}
\email{huhaichuan2024@gmail.com}
\orcid{0009-0002-3007-488X}
\affiliation{
\institution{School of Computer Science and Engineering, Nanjing University of Science and Technology}
\city{Nanjing}\country{China}
}

\author{Chunrong Fang}
\email{fangchunrong@nju.edu.cn}
\orcid{0000-0002-9930-7111}

\author{Zhenyu Chen}
\email{zychen@nju.edu.cn}
\orcid{0000-0002-9592-7022}
\affiliation{
\institution{State Key Laboratory for Novel Software Technology, Nanjing University}
\city{Nanjing}\country{China}
}

\author{Liang Xiao}
\email{xiaoliang@mail.njust.edu.cn}
\orcid{0000-0003-0178-9384}
\affiliation{
\institution{School of Computer Science and Engineering, Nanjing University of Science and Technology}
\city{Nanjing}\country{China}
}

\begin{abstract}
Reproduction tests help developers confirm reported issues and provide executable feedback for issue resolution, yet issue reports in open-source projects rarely include such tests. 
Recent studies have explored generating issue reproduction tests from issue reports with large language models, but existing approaches largely rely on prompt-based pipelines that retrieve textual context and generate tests. 
This limits their ability to understand how reported issues behave in repository-scale codebases and to flexibly organize the construction of reproduction tests.
In this paper, we propose \toolname{}, a multi-stage agent framework for reproduction test generation from issue reports. 
Inspired by how developers manually reproduce reported issues, \toolname{} decomposes the task into four agen stages: bug localization, root cause analysis, test planning, and test generation.
To support these stages, \toolname{} integrates task-specific tools for task decomposition and reflection, context retrieval from both textual sources and repository graphs, and runtime interaction with the execution environment. 
Experiments on \textsc{SWT-bench-lite} and \textsc{SWT-bench-verified} show that \toolname{} successfully reproduces 58.43\% and 70.30\% of issues, outperforming all baselines, with an average cost of \$0.14 per instance.
For example, when equipped with GPT-5-mini, \toolname{} exceeds OpenHands with the same backbone by 20.43 and 7.90 percentage points, respectively.
\toolname{} also generalizes across multiple backbone LLMs and improves downstream issue resolution performance when integrated with existing repair approaches. 
\end{abstract}




\maketitle

\section{Introduction}
\label{sec:intro}

Issue reports are the primary mechanism for submitting bugs on open-source platforms (e.g., GitHub), yet they rarely include executable tests that reproduce the reported issues~\cite{7962388}. 
The absence of such tests makes it difficult for developers to confirm issues before fixing and to validate patches afterward, thereby significantly increasing the cost of software debugging and maintenance~\cite{zhang2023gamma,zhang2024systematic}.
Although software testing has been extensively studied, most existing test generation techniques focus on objectives such as coverage maximization and defect detection~\cite{unittestingSurvey,zhang2025large}. 
In contrast, relatively little attention has been paid to generating tests that faithfully reproduce real-world issues described in natural language, which often involve complex execution contexts and implicit assumptions.

The importance of issue reproduction has been increasingly recognized in recent work, as exemplified by benchmarks such as \textsc{SWT-Bench}~\cite{swtbench}. 
Unlike prior benchmarks that focus on automated program repair~\cite{swebench}, \textsc{SWT-Bench} is specifically designed to evaluate the ability of models to generate tests that reproduce real-world issues from natural language descriptions.
Each instance is constructed from a real-world pull request and consists of an issue description, the corresponding pre-fix codebase, and a reference patch. 
The task is to generate fail-to-pass tests, i.e., tests that fail on the original buggy version but pass after applying the ground-truth fix, thereby serving as executable specifications of the reported issue. 
Such reproduction tests can further facilitate downstream tasks, including but not limited to patch validation and automated repair. 
These benchmarks underscore the critical role of issue reproduction test in verifying patch correctness and enabling iterative repair workflows~\cite{sweagent,openhands}.

Recently, the community has witnessed growing interest in automatically generating issue reproduction tests from issue reports using Large Language Models (LLMs)~\cite{zhang2026survey}.
Representative approaches include \textsc{LIBRO}~\cite{libro}, which leverages few-shot prompting with example issue–test pairs, and \textsc{Issue2Test}~\cite{issue2test}, which incorporates meta-prompting to infer project-specific testing conventions followed by iterative refinement based on execution feedback. 
\textsc{AssertFlip}~\cite{assertFlip} adopts a different strategy by first generating passing tests and then inverting assertions, based on the observation that LLMs are more effective at producing valid passing tests.
Despite these advances, existing prompt-based approaches largely follow a similar pipeline of context retrieval and test generation, which introduces two fundamental limitations.
First, reproducing an issue often requires understanding behavior along execution paths that span multiple modules and files, whereas existing approaches primarily rely on text-based retrieval, making it difficult to capture the cross-file behavioral dependencies involved in the reported issue. 
Second, generating tests is inherently a multi-step task that requires bug localization, root cause analysis, test setup construction, and test verification, whereas existing approaches largely treat it as direct generation from retrieved context, without explicitly modeling the intermediate planning process.

To address these challenges, we propose \toolname{}, a multi-stage agent framework for issue reproduction test generation. 
Rather than treating reproduction test generation as a fixed prompting pipeline, \toolname{} decomposes the task into multiple agent stages and equips each stage with task-specific tools for autonomous exploration and decision-making. 
This design enables \toolname{} to support both repository-scale issue understanding and the structured construction of reproduction tests. 
In particular, \toolname{} integrates three categories of tools: task decomposition and reflection tools for progressively reasoning, code retrieval tools for collecting issue-relevant context from textual sources and repository graph, and runtime interaction tools for interacting with the runtime environment.
Built upon this toolset, \toolname{} organizes the generation process into four stages: bug localization via hierarchical analysis, root cause analysis via execution path, assertion-aware test planning, and test generation via triadic review.
This design is inspired by the workflow developers typically follow when manually reproducing reported issues: they identify issue-relevant code, reason about how the failure is triggered, design the test setup and assertions, and execute the test to check whether the observed failure matches the report.

Experiments on \textsc{SWT-bench-lite} and \textsc{SWT-bench-verified} demonstrate that \toolname{} successfully reproduces 58.43\% and 70.30\% of issues, outperforming all baselines, e.g., improving \textsc{AssertFlip} by 53.76\% and 54.51\%. 
\toolname{} also generalizes across multiple backbone LLMs, achieving average rates of 64.37\%, 58.47\%, 52.11\%, and 50.43\% with GPT-5-mini, Qwen3-Coder, DeepSeek-V3.2, and GLM-4.6, respectively.
Moreover, additional discussions demonstrate that \toolname{} can be integrated with existing repair approaches, increasing the number of successfully resolved issues from 143 to 153 for \textsc{SWE-agent} on  \textsc{SWT-bench-lite}.

In summary, this paper makes the following contributions:
\begin{itemize}
    \item \textbf{Multi-stage Agents.} We propose \toolname{}, a multi-stage agent framework for repository-level issue reproduction test generation, which decomposes the task into bug localization, root cause analysis, test planning, and test generation.
    
    \item \textbf{Task-specific Tools.} 
    We design a dedicated toolset to support this framework, including task decomposition and reflection tools, context retrieval tools, and runtime interaction tools.
    
    \item \textbf{Extensive Evaluation.} We conduct extensive experiments on two benchmark datasets against several state-of-the-art baselines.
    The results show that \toolname{} achieves significant improvements. Ablation studies validate the effectiveness of individual components, and integration with existing repair frameworks further demonstrates its practical value.

\end{itemize}

\section{Background and Motivation}
\label{sec:background}

\subsection{Test Generation}

Automated test generation has been extensively studied in software engineering~\cite{unittestingSurvey}. Existing approaches can be broadly grouped into three paradigms: traditional, learning-based, and LLM-based.

\textbf{Traditional Approaches}.
These studies are mainly represented by heuristic search and random testing techniques~\cite{korel1990automated,meyer2007autotest,fraser2013whole,mcdminosa2015mosa,panichella2018dynamosa}.
Representative examples include EvoSuite~\cite{evosuite}, which uses genetic algorithms to evolve test suites toward predefined coverage objectives, and Randoop~\cite{randoop}, which generates method invocation sequences through feedback-directed random testing. 
These approaches are effective at improving structural coverage, although prior studies have shown that high coverage does not necessarily imply strong bug detection capability~\cite{testCoverage,chekam2017mutation}.

\textbf{Learning-based approaches}.
These studies typically formulate test generation as a sequence-to-sequence learning problem over given focal methods. 
For example, AthenaTest~\cite{athenatest} constructs large-scale focal method--test pairs and reports strong performance on Defects4J~\cite{defects4j}.
\textsc{A3Test}~\cite{a3test} further improves assertion correctness through knowledge injection and test validity verification.

\textbf{LLM-based approaches}.
Recent studies have shown that LLM can generate unit tests with promising effectiveness, while also motivating a shift from direct one-shot prompting to more structured generation pipelines~\cite{chatgpt4unittest,wang2024hits,ryan2024code,yin2025enhancing,jain2025testgeneval,zhang2025testbench,cheng2025rug,gu2024testart}.
Representative approaches further improve this paradigm from different angles.
ChatUniTest~\cite{chen2024chatunitest} enhances generation with adaptive focal-context construction and automated validation,
CoverUp~\cite{altmayerpizzorno2025coverup} introduces coverage-guided iterative refinement,
and IntUT~\cite{nan2025test} improves generation by explicitly modeling test intentions such as inputs, mocks, and expected outcomes.

Despite their effectiveness, these approaches are not tailored to reproduction test generation.
Existing approaches typically assume a given testing target and a functionally correct implementation, with the goal of improving coverage or validating expected behavior. 
By contrast, our work starts from a natural-language issue report over a buggy repository, and the objective is to generate a failing test that reproduces the reported issue. 
This setting requires understanding issue semantics, localizing relevant elements, and determining whether the resulting failure truly corresponds to the described issue, making the task substantially challenging.

\subsection{Bug Reproduction Test Generation}

As a special form of test generation, this task focuses on constructing tests to reproduce issues described in software issue trackers. 
Existing work can be discussed from three perspectives: benchmarks and task-specific approaches.

\textbf{Benchmarks.}
SWE-bench~\cite{swebench} is a a representative early benchmark to evaluate the ability of LLMs to resolve real-world issue reports from software repositories.
It is built from real-world issues collected from 12 Python repositories and has become a widely used testbed for repository-level software engineering tasks.
Built on top of SWE-bench, SWT-bench~\cite{swtbench} focuses specifically on bug reproduction test generation. It includes two subsets, SWT-bench-lite and SWT-bench-verified, where the latter is a manually curated subset with verified correctness. These benchmarks provide a common evaluation basis for studying reproduction test generation in realistic repository settings.

\textbf{Task-specific Approaches.}
These approaches mainly use LLMs to transform issue descriptions into executable reproduction tests through prompting and feedback-based refinement~\cite{wang2025aegis,wang2026icore,hora2026brt,ahmed2025otter,fei2026echo}.
For example, \textsc{LIBRO}~\cite{libro} is among the earliest studies in this direction.
It adopts few-shot prompting with issue-test pairs, followed by post-processing and reranking to improve generation quality, but makes limited use of repository-specific context.
\textsc{Issue2Test}~\cite{issue2test} further incorporates root-cause analysis, meta-prompting for project-specific testing conventions, and iterative refinement with execution feedback.
\textsc{AssertFlip}~\cite{assertFlip} explores a complementary strategy.
It first generates passing tests that capture the buggy behavior, and then flips assertions to obtain bug-reproducing tests.
However, most existing methods still rely on fixed prompting pipelines, primarily textual context construction, or execution feedback without explicit semantic review.
Although some methods, such as \textsc{Issue2Test}, incorporate root-cause analysis, and execution feedback, these capabilities are mostly organized as predefined pipeline steps rather than as a task-specific agentic loop.
In contrast, \toolname{} organizes reproduction test generation as a multi-stage agentic process that explicitly integrates bug localization, root-cause analysis, assertion-aware test planning, and triadic review with execution feedback.

\subsection{General SE Agents}
More recent SE agents are designed for broad repository-level software engineering tasks, such as issue resolution~\cite{zhang2026sgagent}.
Within these general-purpose workflows, reproduction test generation is often incorporated as an intermediate step for understanding the issue or validating candidate patches.
For example, OpenHands~\cite{openhands} employs ReAct-style agents~\cite{yao2023react} to navigate repositories and construct verification tests.
SWE-agent~\cite{sweagent} defines an issue-resolution workflow in which bug localization is followed by reproduction test generation.
Agentless~\cite{agentless} further reformulates repository-level repair as a fixed workflow, including fault localization, patch generation, and filtering with regression and reproduction tests.
These systems demonstrate the value of agentic or workflow-based reasoning for repository-level software engineering.
However, because their primary goal is issue resolution rather than faithful reproduction test generation, reproduction tests are usually treated as auxiliary artifacts.
As a result, they place limited emphasis on systematically verifying whether the generated tests faithfully reproduce the reported bug.

\begin{figure}[t]
    \centering
    \includegraphics[width=0.9\linewidth]{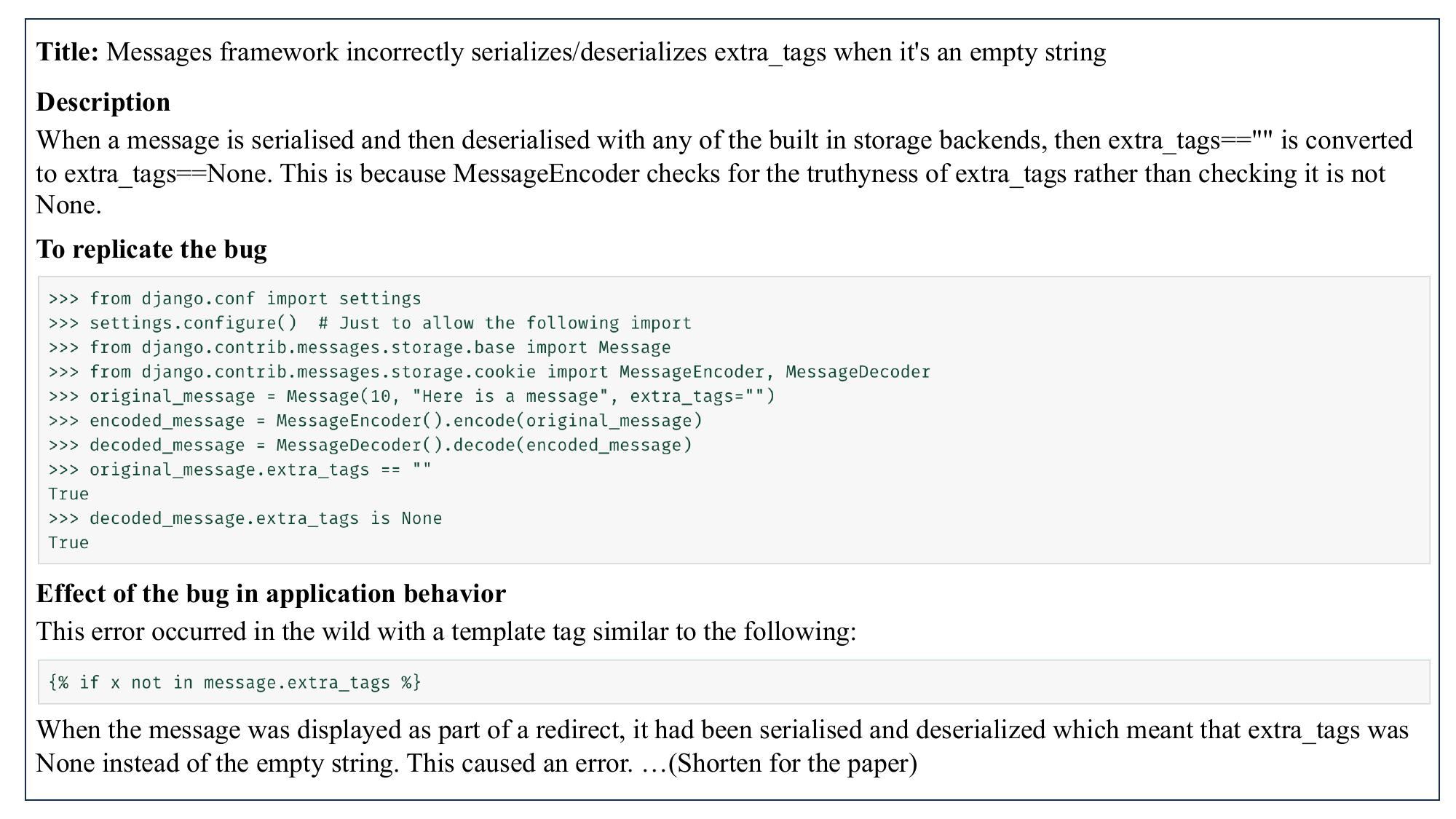}
    \caption{The django-15347 issue from SWT-bench}
    \label{fig:case}
\end{figure}

\subsection{Motivating Example}
\label{section:case-study}

We use a real-world issue \texttt{django-15347} to illustrate the challenges in repository-level issue reproduction test generation.
As shown in \autoref{fig:case}, the issue concerns the handling of the \texttt{extra\_tags} field in Django's messages framework. When a message with the field \texttt{extra\_tags=""} is serialized and deserialized, the resulting object incorrectly sets \texttt{extra\_tags} to \texttt{None}. Although the issue appears simple from its description, reproducing it requires reasoning about the semantic difference between an empty string and a missing field, as well as tracing how this discrepancy propagates through the underlying serialization and deserialization logic.

\begin{lstlisting}[float=t,language=Python, caption={Test generated by \toolname{} for Django-15347}, label={lst:django-15347-test}, basicstyle=\footnotesize\ttfamily, breaklines=true, frame=single]
import pytest
from django.conf import settings
from django.contrib.messages.storage.base import Message
from django.contrib.messages.storage.cookie import MessageEncoder, MessageDecoder

def test_reproduce_issue():
    if not settings.configured:
        settings.configure()

    original_message = Message(10, "Here is a message", extra_tags="")
    encoder = MessageEncoder()
    encoded_message = encoder.encode(original_message)
    decoder = MessageDecoder()
    decoded_message = decoder.decode(encoded_message)

    assert original_message.extra_tags == "", "Original extra_tags should be empty string"
    assert decoded_message.extra_tags == "", "Decoded extra_tags should be empty string"
    assert original_message.extra_tags == decoded_message.extra_tags, "extra_tags should be preserved"
\end{lstlisting}

This issue is challenging because the issue report does not explicitly reveal the faulty code location or the bug-triggering execution path. 
To reproduce it correctly, the model must identify the relevant repository context, connect the serialization and deserialization logic across functions, and construct an assertion that captures the semantic mismatch between \texttt{""} and \texttt{None}. 
In our experiments, both \textsc{LIBRO} and \textsc{Issue2Test} fail to generate a correct reproduction test for this issue. 
This failure highlights two core challenges in reproduction test generation: understanding repository-scale issues and constructing generation stages.
\toolname{} is designed to address these challenges through a multi-stage process that combines bug localization, root cause analysis, test planning, and test generation.
For this issue, \toolname{} successfully generates the reproduction test shown in Listing~\ref{lst:django-15347-test}, which creates a \texttt{Message} object with \texttt{extra\_tags=""}, serializes and deserializes it, and asserts that the empty string should be preserved rather than converted to \texttt{None}. This example shows why effective issue reproduction test generation requires repository exploration, semantic issue understanding, and failure-aware validation rather than test generation alone.

\section{Approach}
\label{sec:approach}

\subsection{Overview}

\begin{figure*}[t]
    \centering
    \includegraphics[width=0.9\textwidth]{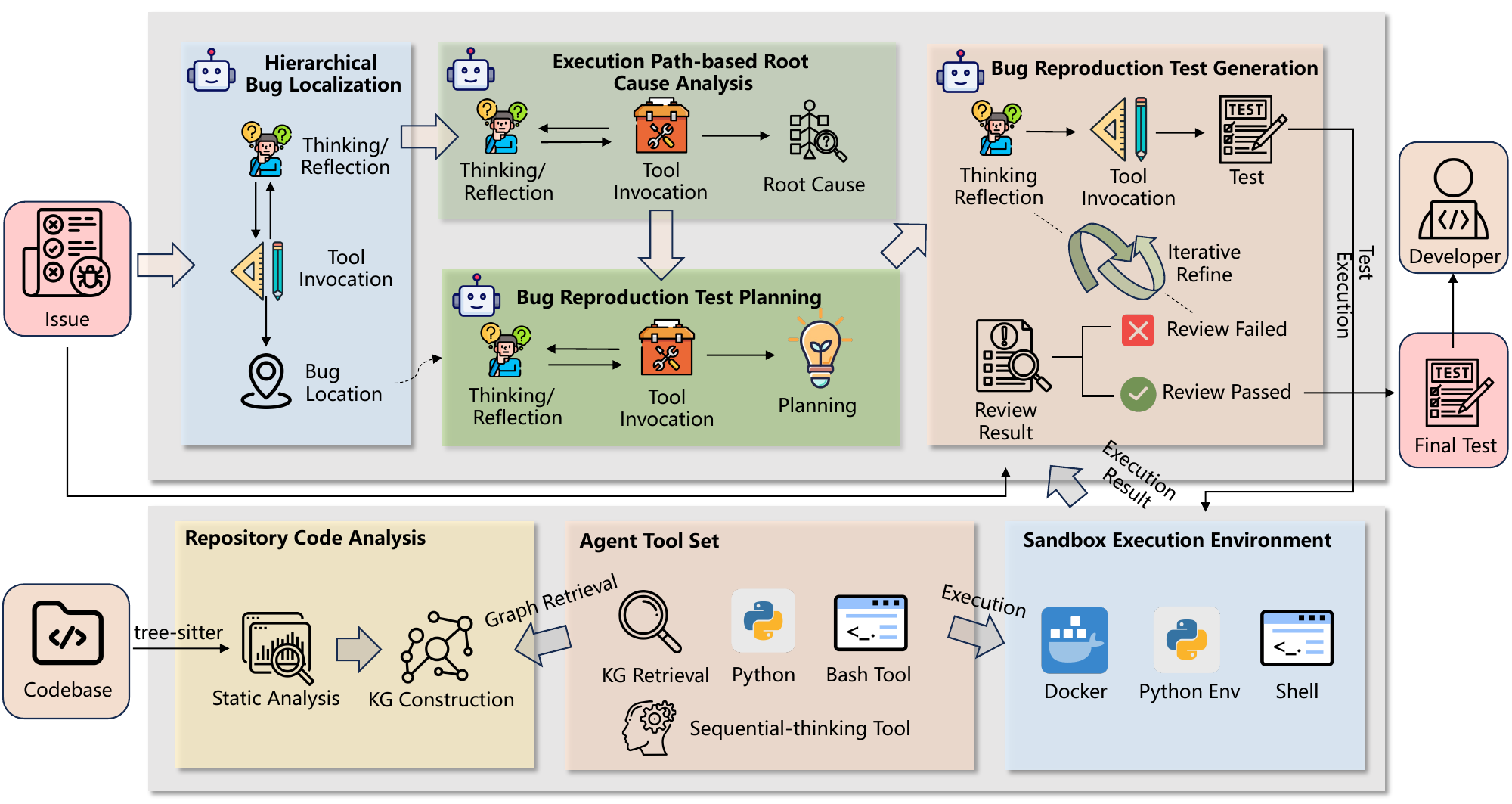}
    \caption{Overview of the \toolname{} framework}
    \label{fig:framework}
\end{figure*}

As illustrated in \autoref{fig:framework}, \toolname{} is designed as a multi-stage agent framework for repository-level issue reproduction test generation. 
Particularly, \toolname{} equips LLMs with three categories of tools: task decomposition and reflection tools, hybrid context retrieval tools, and runtime interaction tools. 
These tools allow the agent to break down complex objectives, capture relevant code and dependencies, interact with the repository environment, and refine its decisions based on intermediate feedback.

Built on top of this toolset and inspired by how developers manually reproduce reported issues, \toolname{} formulates issue reproduction test generation as a four-stage agentic workflow: bug localization via hierarchical analysis, root cause analysis via execution path, assertion-aware test planning, and test generation via triadic review.
In the bug localization stage, the agent identifies suspicious files, classes, functions, and code regions related to the reported issue. In the root cause analysis stage, it analyzes relevant implementations, call paths, and dependencies to infer the bug-triggering execution path and underlying cause. In the test planning stage, it examines existing tests and project-specific testing conventions to derive an appropriate reproduction strategy, including setup, invocation patterns, and expected failure conditions. 
In the test generation stage, it generates candidate reproduction tests, executes them in a Docker-based sandbox, and iteratively refines them based on execution feedback and triadic review.

\subsection{Agent Toolset Construction}
Issue reproduction test generation over repository-level codebases is inherently challenging, as it requires understanding complex implementations, cross-file dependencies, and dynamic execution behavior. 
Unlike prior approaches that largely follow a fixed prompting pipeline, \toolname{} adopts a multi-stage agent framework that performs iterative exploration and decision-making throughout the process. 
To support this framework, we equip the agent with a multi-dimensional toolset at each stage, enabling it to decompose tasks, retrieve relevant context, interact with the repository environment, and refine its decisions based on intermediate feedback.

As shown in Table~\ref{table:tools}, the toolset consists of three categories: task decomposition and reflection tools, hybrid context retrieval tools, and runtime interaction tools. 
These tools provide complementary support across the pipeline. 
Task decomposition and reflection tools help the agent break down complex objectives and revise intermediate reasoning. 
Context retrieval tools enable the agent to inspect relevant code and identify dependencies among modules, classes, and functions. 
Runtime interaction tools allow the agent to observe execution behavior and validate hypotheses in the actual repository environment.

\begin{table*}[htbp]
    \centering
    \footnotesize
    \caption{Equipped Toolset in \toolname{}}
    \resizebox{\textwidth}{!}{
    \begin{tabular}{lll}
        \toprule
        Tool Name & Tool Category & Function Description \\
        \midrule
        Sequential-Thinking    
            & Task Decomposition and Reflection Tool   
            & Decomposes complex tasks and supports process-level reflection  \\ 
        \midrule
        Grep      
            & Context Retrieval Tool  
            & Searches file contents based on keywords  \\
        Glob  
            & Context Retrieval Tool  
            & Matches file names based on keywords  \\
        Read-File
            & Context Retrieval Tool  
            & Reads file contents with optional range selection  \\
        Graph-Search 
            & Context Retrieval Tool
            & Retrieves dependency relations based on a code knowledge graph  \\
        \midrule
        Bash 
            & Runtime Interaction Tool  
            & Executes command-line operations  \\
        Interactive-Python 
            & Runtime Interaction Tool     
            & Executes interactive Python scripts at runtime  \\
        \bottomrule
    \end{tabular}
    }
    \label{table:tools}
\end{table*}

\subsubsection{Task Decomposition and Reflection Tools}

Issue reproduction test generation requires multi-stage reasoning, long-horizon planning, and iterative adaptation. Unlike prior approaches that follow a fixed pipeline, \toolname{} adopts a multi-stage agent framework in which each stage performs fine-grained task decomposition and dynamically adjusts its actions according to intermediate results and environmental feedback. To support this process, we incorporate sequential-thinking~\cite{SequentialThinking,ModelContextProtocol_Specification} as an explicit task decomposition and reflection tool. It allows the agent to break down each stage into manageable reasoning steps, revise earlier decisions when necessary, and adapt its strategy as new evidence becomes available.

Within \toolname{}, sequential-thinking serves as a high-level controller for structured decomposition and dynamic reflection. At each step, the agent maintains an explicit reasoning trajectory with step indices and estimated total steps, while using explicit control signals to determine whether to continue, revise previous thoughts, or branch into alternative reasoning paths. Concrete actions such as context retrieval, execution, and test generation are delegated to other tools, and their outputs are fed back into subsequent reasoning steps. This design forms a closed loop of \emph{thinking--acting--observing--revising}, which improves the stability and interpretability of long-horizon reasoning.

\subsubsection{Hybrid Context Retrieval Tools}
\toolname{} provides two complementary retrieval mechanisms: command-line keyword-based retrieval and knowledge-graph-based dependency retrieval.

\textbf{Command-line keyword-based retrieval.}
\toolname{} provides Linux-style retrieval tools that mimic how developers explore repositories from the command line. These tools support repository navigation, file localization, keyword lookup, and line-level source inspection through operations analogous to \texttt{ls}, \texttt{find}, \texttt{grep}, and \texttt{glob}. They allow the agent to quickly build an understanding of repository layout, module organization, and implementation details relevant to the reported issue. To avoid wasting the LLM context window, excessively long outputs are truncated or summarized. In addition, potentially dangerous commands are filtered before execution, and all tool calls are executed inside a Docker-based sandbox to ensure safety.

\textbf{Knowledge-graph-based dependency retrieval.}
To support deeper semantic understanding and cross-file reasoning, \toolname{} also incorporates a code knowledge graph. The graph is constructed by recursively parsing the repository with Tree-sitter~\cite{tree_sitter} to extract program entities such as classes, function definitions, and function calls from source files. These entities are represented as nodes, while relations such as file membership, class references, and function calls are encoded as edges, forming a cross-file semantic dependency network.
Built on top of this graph, \toolname{} provides dependency-oriented retrieval tools such as entity lookup, class structure inspection, cross-file relation querying, and import analysis. 
Unlike keyword-based retrieval, these tools directly exploit semantic dependencies and return structurally relevant context, improving the agent’s ability to recover call chains, dependency paths, and behaviorally related code regions. 

\subsubsection{Runtime Interaction Tools}

While context retrieval tools focus on static repository information, runtime interaction tools expose the repository's dynamic behavior during execution. Their purpose is to enable the agent to validate hypotheses, inspect runtime behavior, and revise its decisions based on concrete execution feedback.

To support this process, \toolname{} provides two types of runtime interaction tools. The first is a  command execution tool, which allows the agent to run repository-level commands inside a Docker-based sandbox. It supports environment inspection, dependency checking, build execution, and test invocation. The second is an interactive  execution tool, which allows the agent to execute Python snippets directly within the sandboxed repository. This tool enables fine-grained inspection of function behavior, dependency interactions, and data-flow changes under concrete inputs.

\subsection{Bug Localization via Hierarchical Analysis}

Given an issue report and a target repository, this module attempts to identify a ranked set of suspicious program elements that are likely related to the reported failure.
To this end, we design a multi-stage hierarchical analysis procedure that progressively narrows the search space from suspicious files to fine-grained code lines. 
At the core of this procedure is a chain-of-thought-style reasoning process, in which the agent incrementally forms, refines, and verifies localization hypotheses based on evidence collected from the issue report and the repository context.

Starting with the issue report, \toolname{} first constructs an initial query by extracting key symptoms, error messages, and referenced entities, such as file names, class names, and method names from stack traces. 
Guided by this initial query,  \toolname{} gathers contextual evidence through syntactic and semantic retrieval mechanisms. 
Specifically, it uses command-line-based text retrieval to search for issue-related files using keywords such as exception names, function names, log fragments, line numbers, and configuration parameters. 
In parallel, it performs knowledge-graph-based dependency retrieval to analyze call chains across files and inter-module dependency structures, thereby inferring how faults may propagate through the system.
After identifying suspicious files associated with the issue, \toolname{} reads the relevant implementations using file-reading tools for issue understanding. 
The retrieved contexts are then iteratively refined through a reasoning-and-retrieval loop, where intermediate hypotheses about potential root causes are generated and used to reformulate queries for subsequent retrieval steps.

Once candidate files has been identified, \toolname{} performs hierarchical localization within each file. It first ranks functions or classes according to their semantic relevance to the issue and their structural proximity to previously identified elements, such as those connected through call relations or dependency links.
It then further refines the search to the line level by examining code snippets, execution-relevant statements, and context-specific cues, including conditionals, error-handling logic, and data-flow usage. 
This coarse-to-fine localization strategy enables efficient exploration of large codebases while maintaining localization precision.

To further encourage explicit reasoning, \toolname{} requires the agent to provide a natural-language rationale for each selected suspicious region, together with a confidence score. These rationales externalize the agent's reasoning trajectory and make the localization process more interpretable, while the confidence scores offer an additional signal for prioritizing candidates in downstream stages.
Finally, the module outputs a ranked list of suspicious locations, each accompanied by a confidence estimate and supporting rationale.
These results are then passed to the downstream root cause analysis and test generation stages.

\subsection{Root Cause Analysis via Execution Path}

After bug localization, \toolname{} performs root cause analysis to infer the failure mechanism behind the reported issue.
Given the suspicious locations from the previous stage and the issue description, this stage traces how the fault is triggered and propagated along the execution path, and produces a structured root cause analysis report for downstream issue reproduction test generation.

\toolname{} first analyzes the issue report together with the suspicious locations to identify the reported failure symptoms, such as exception types, error messages, incorrect return values, and abnormal state changes. 
Based on these symptoms, the agent then performs execution path analysis by incrementally retrieving relevant program context, including dependent functions, related classes, and other critical code regions, and reconstructing the execution path from the entry point to the failure point.
After reconstructing the execution path, the agent identifies the underlying logical defect. This step focuses on determining what is wrong in the code logic and why it leads to the unexpected behavior described in the issue report. By comparing the intended behavior implied by the issue report with the actual behavior exhibited by the code, the agent identifies common bug patterns such as incorrect condition checks and improper boundary handling.

\begin{figure}[!htb]
    \centering
    \includegraphics[width=0.7\textwidth]{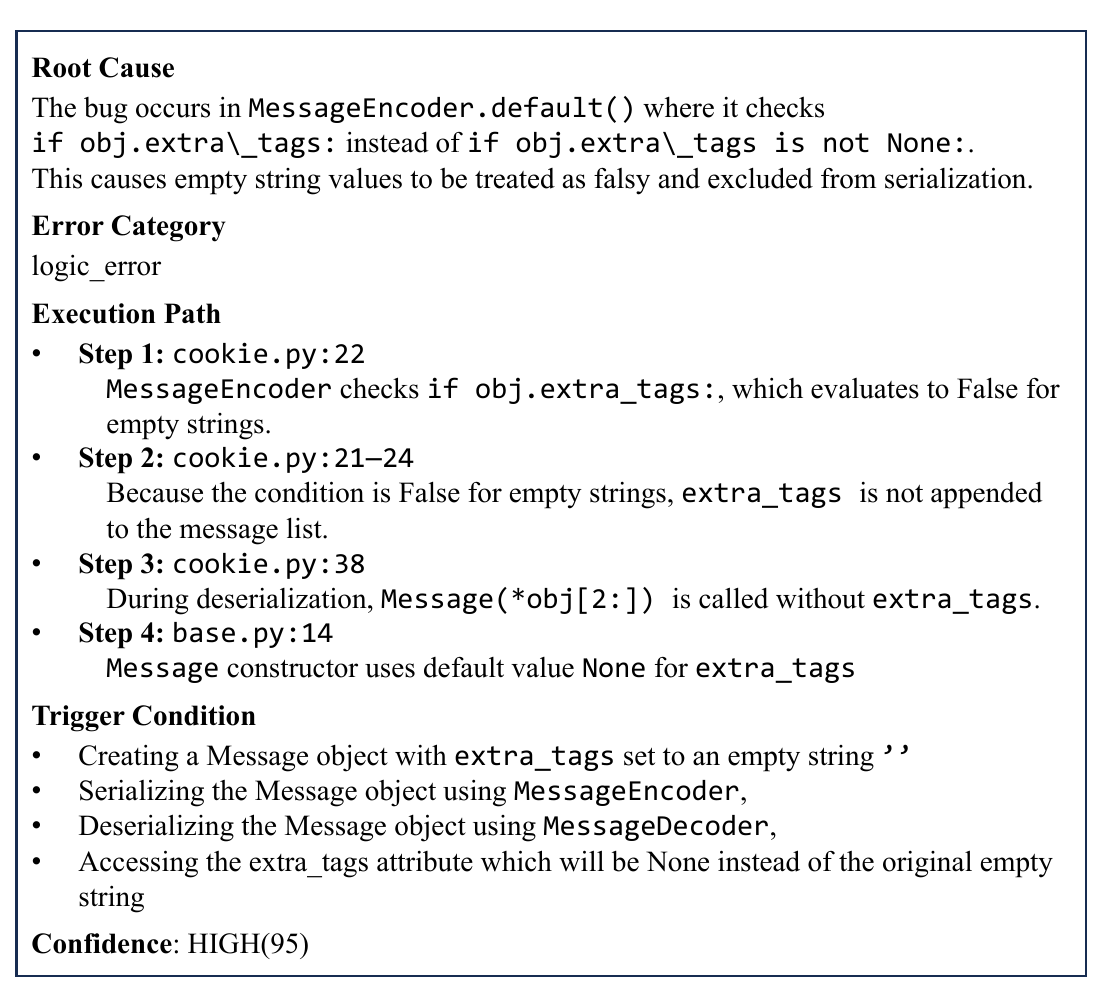}
    \caption{Root Cause Analysis result of Django-15347}
    \label{fig:case_root}
\end{figure}

The agent then formulates a root cause hypothesis based on symptom interpretation, execution path analysis, and logical defect inspection. This hypothesis provides an abstract explanation of the failure mechanism: it not only identifies where the defect lies, but also explains why that defect produces the reported symptoms. To improve reliability, \toolname{} further refines this hypothesis through an iterative reason--validate--refine loop. During this process, the agent continuously collects supporting evidence, including relevant code snippets, complete method implementations, call dependencies, and module relationships, and revises the hypothesis until it consistently explains both the observed symptoms and the retrieved program context.
Finally, as shown in \autoref{fig:case_root}, \toolname{} outputs a structured root cause analysis report with five components: (1) root cause summary, a concise description of the underlying fault mechanism; (2) error category, the defect type, such as logic error, boundary-handling defect, or API misuse; (3) execution path, the key execution steps from the entry point to the failure point, annotated with corresponding code locations and behaviors;
(4) trigger conditions, the environmental constraints required to reproduce the failure; and (5) confidence score, a measure of the reliability of the inferred root cause.

\subsection{Assertion-aware Test Planning}

After root cause analysis, \toolname{} performs test planning to derive structured guidance for downstream test generation.
Given the root cause analysis report, this stage identifies the trigger conditions, reproduction setup, and assertions to reproduce the reported issue, thereby reducing structural errors and information omissions that often arise in direct test generation.
Specifically, this stage consists of three connected steps: trigger condition analysis, reproduction setup construction, and assertion derivation.

In trigger condition analysis, \toolname{} first parses the root cause analysis report to extract key information, such as the root cause summary, issue category, and execution path.
Based on this information, it derives the minimal trigger conditions and key constraints required to expose the issue. 
In reproduction setup construction, \toolname{} determines the preconditions and environmental constraints required for reproduction, such as project initialization procedures, dependency loading mechanisms, necessary context states, and version-specific requirements.
It further identifies the minimal test inputs, parameter combinations, and boundary values sufficient to reproduce the issue, using retrieval and runtime interaction tools to ground the plan in the actual repository context.
In assertion derivation, \toolname{} determines how the reproduced bug should be verified. It analyzes observable failure signals and runtime outputs that characterize the issue behavior, such as exception messages, stack traces, incorrect return values, and abnormal internal states. 
Based on these observations, \toolname{} specifies the expected behavior differences between buggy and intended execution, and determines the assertions that should be encoded in the reproduction test.
Finally, \toolname{} outputs a structured test plan with three components: 
(1) environment constraints, which specify the setup and initial states for reproduction; 
(2) test input construction, which defines the minimal inputs and key parameter configurations needed to trigger the bug; 
and (3) expected assertion, which describes the deviations between buggy and intended behavior and specifies how they should be asserted.

\subsection{Test Generation via Triadic Review}

After test planning, \toolname{} enters the test generation stage. This stage takes as input the intermediate artifacts produced in the previous stages, including the issue description, suspicious locations, root cause analysis report, and test plan, and outputs an executable issue reproduction test together with its execution result. Its goal is to translate the structured analysis results into a runnable test and, through iterative execution and review, ensure that the generated test can reliably reproduce the reported issue.

To generate the reproduction test, \toolname{} assembles the issue description, suspicious locations, root cause analysis report, and test plan into a unified prompt for the agent. Guided by these inputs, the agent constructs a single reproduction test that targets the relevant suspicious code regions, instantiates the necessary setup and input conditions for issue manifestation, and encodes assertions against the expected faulty behavior. The generated test is required to expose the actual issue behavior, that is, it should fail on the buggy version and pass once the issue is fixed.
Once a candidate test is generated, \toolname{} writes it into the sandboxed repository and executes it in a Docker-based environment. 
The execution outputs are then used as feedback for subsequent refinement. 
If the test passes, it fails to reproduce the issue, and the agent must revise the test based on the observed execution feedback. A failing test, however, is not immediately accepted as a valid reproduction. Instead, \toolname{} further examines whether the observed failure is semantically consistent with the reported issue. Only tests whose failures are confirmed to match the issue description are accepted as valid reproduction tests; otherwise, the agent continues to refine the test through further iterations until the review succeeds or the maximum iteration budget is reached.

\begin{figure}[!htb]
    \centering
    \includegraphics[width=0.8\linewidth]{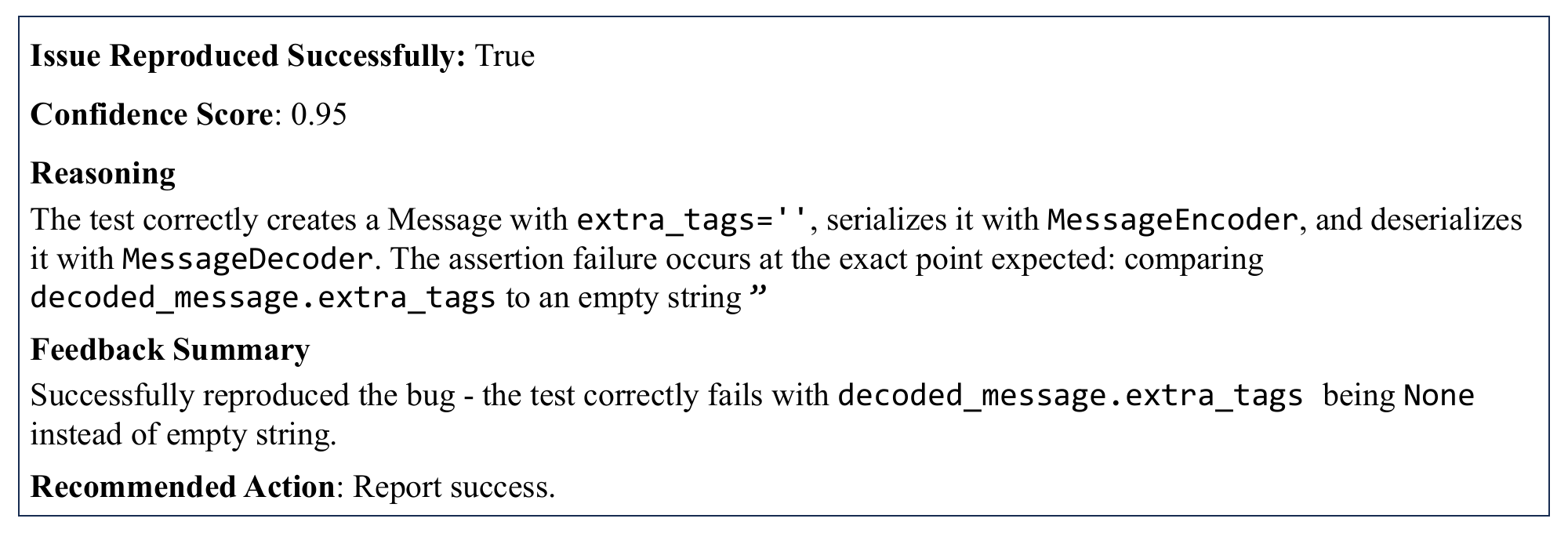}
    \caption{Review result of Django-15347}
    \label{fig:case_review}
\end{figure}

To determine whether a failing test truly reproduces the reported issue, we introduce a triadic review procedure over the issue report, the generated test, and the execution result, as shown in \autoref{fig:case_review}.
The procedure evaluates: (1) whether the generated test is semantically aligned with the issue report in its targeted functionality, triggering conditions, and assertions; (2) whether the observed failure is consistent with the test logic rather than caused by incidental factors such as syntax errors, missing dependencies, or environment misconfiguration; and (3) whether the execution result matches the issue report in terms of error message, exception type, or overall failure pattern.
If a generated test passes review, it is retained as a valid issue reproduction test and can be used by developers for inspection, validation, and regression checking. Otherwise, the agent uses the review result together with the prior execution feedback and intermediate analysis artifacts to further refine the test, including its preconditions, input construction, and assertions, and then re-executes the revised version. 
In this way, \toolname{} forms a closed-loop framework based on feedback-driven iteration and triadic review, enabling the agent to progressively converge to a high-quality test that reliably reproduces the target issue.

\section{Experimental Setup}
\label{sec:experiment}

\subsection{Research Questions}

To evaluate the  effectiveness of \toolname{}, we conduct experiments to answer the following research questions (RQs):

\begin{itemize}[leftmargin=*]
    \item \textbf{RQ1:} How does \toolname{} perform compared to state-of-the-art approaches?
    \item \textbf{RQ2:} What is the impact of different LLMs on the generalizability of \toolname{}?
    \item \textbf{RQ3:} What are the contributions of different components in \toolname{}?
\end{itemize}

\subsection{Datasets}
\label{sec:datasets}

We evaluate \toolname{} on two subsets from \textsc{SWT‑Bench}~\cite{swtbench}: \textsc{SWT‑Bench‑Lite} and \textsc{SWT‑Bench‑Verified}. 
\textsc{SWT‑Bench} is a large‑scale benchmark for evaluating the capability of models and code agents to generate reproducible tests for real‑world issues collected from GitHub repositories. 
\textsc{SWT‑Bench‑Verified} is a curated subset comprising human‑verified tasks with clear descriptions and well‑specified reproduction criteria, providing a higher‑confidence evaluation set. 
\textsc{SWT‑Bench‑Lite} is a smaller, lightweight subset designed for rapid iteration and efficient benchmarking. 
Both subsets have been widely adopted in recent work on reproducible test generation for real‑world issues~\cite{otter,issue2test,assertFlip,libro}.

\subsection{Baselines}
We compare \toolname{} with several representative baselines from two categories.

First, we include task-specific reproduction-test generation methods.
Zero-Shot-Plus~\cite{swtbench} is a zero-shot prompting baseline introduced by \textsc{SWT-Bench}.
\textsc{LIBRO}~\cite{libro} adopts few-shot prompting with issue--test pairs, followed by post-processing and reranking to select high-quality candidate tests.
\textsc{Issue2Test}~\cite{issue2test} employs a multi-stage pipeline with root-cause analysis, meta-prompting for project-specific testing conventions, and iterative refinement with execution feedback.
\textsc{Otter}~\cite{otter} performs bug localization and uses reflection-based planning to iteratively generate and refine tests.
\textsc{Otter++}~\cite{otter} extends Otter with heterogeneous prompt sampling and ensemble-style selection.
\textsc{e-Otter} and \textsc{e-Otter++}~\cite{eotterpp} further incorporate execution-feedback-guided generation and selection.
\textsc{AssertFlip}~\cite{assertFlip} first generates passing tests that capture buggy behavior and then inverts their assertions to obtain bug-reproducing tests.

Second, we include general-purpose software engineering agents.
AutoCodeRover~\cite{authCodeRover} is an LLM-based issue repair framework adapted in \textsc{SWT-Bench} to produce bug reproduction tests through modified instructions.
\textsc{SWE-Agent}~\cite{sweagent}, \textsc{SWE-Agent+}~\cite{swtbench}, \textsc{Aider}~\cite{aider_github}, and OpenHands~\cite{openhands} are coding agents that can be adapted to generate reproduction tests.
Amazon Q~\cite{aws_amazon_q_2026} is an LLM-based development assistant, and we use its publicly reported results on the \textsc{SWT-Bench} leaderboard.
\subsection{Evaluation Metrics}
\label{sec:metrics}
Following the evaluation protocols widely adopted in prior reproduction test generation work~\cite{swtbench}, we assess the effectiveness of \toolname{} using the Fail-to-Pass (FP) Rate. 
Given one generated test per issue, the FP Rate measures the proportion of tests that both reproduce the reported issue and validate its corresponding patch, i.e., tests that fail on the pre-patch (buggy) version of the code and pass after the issue-resolving patch is applied. 
This metric directly captures whether a generated test correctly characterizes the buggy behavior and serves as a reliable test after the fix.

\subsection{Implementation Details}
To implement \toolname{}, we adopt GPT-5-mini~\cite{openai_gpt5} as the primary backbone model. GPT-5 mini belongs to OpenAI's GPT-5 model family, which is designed to support coding and agentic tasks involving long-context reasoning and tool interaction.
This makes it suitable for our multi-stage workflow, where the agent must localize bugs, analyze execution paths, plan assertions, and iteratively refine reproduction tests based on runtime feedback.
We also instantiate \toolname{} with other advanced code-capable LLMs, including DeepSeek-V3.2, GLM-4.6, and Qwen3-Coder, to evaluate whether the proposed framework generalizes across different backbone models.
The overall framework is built upon LangChain~\cite{langchain} and LangGraph~\cite{langgraph}, which are used to orchestrate multi-stage reasoning and tool interactions. 
We configure the LLM with temperature = 0.7 and top\_p = 0.8; for open-source backbones, we additionally set top\_k = 20 and repetition\_penalty = 1.05 when these decoding parameters are supported.
To control computational cost, we enforce a bounded interaction budget.
Each stage allows up to 50 LLM calls. 
In the test generation stage, which incorporates feedback-driven refinement and automated validation, the maximum number of feedback iterations is limited to 20.
All experiments are conducted on a server running Ubuntu 22.04. 
For knowledge graph construction, we utilize Neo4j~\cite{neo4j}, which enables efficient representation and querying of code structure and dependencies.

\section{Evaluation and Results}
\label{sec:results}

\subsection{RQ1: Comparison with State-of-the-Arts}

\textbf{Experimental Design.}
We evaluate the effectiveness of \toolname{} on two widely used benchmarks, \textsc{SWT-bench-lite} and \textsc{SWT-bench-verified}. 
Following prior SWT-bench studies, we adopt FP Rate as the primary evaluation metric, which measures whether a generated test fails on the buggy version and passes on the patched version. 
We compare \toolname{} with representative prompting-based and agent-based baselines, including \textsc{LIBRO}, \textsc{Issue2Test}, \textsc{Otter}, \textsc{Otter++}, OpenHands, and \textsc{AssertFlip}, as well as additional systems with publicly reported SWT-bench leaderboard results. Unless otherwise specified, the baseline results are taken from the corresponding original papers or the public \textsc{SWT-bench} leaderboard.
To control for backbone-model effects, we further include a same-backbone comparison with the strongest baseline OpenHands under GPT-5-mini. 
Specifically, on \textsc{SWT-bench-lite}, we reproduce OpenHands using the recommended \textsc{SWT-bench} configuration~\cite{swtbench_openhands_config}, while on \textsc{SWT-bench-verified}, we use the result reported on the \textsc{SWT-bench} leaderboard.

\begin{table}[t]
    \centering
    \caption{Comparison of \toolname{} with baselines on \textsc{SWT-bench-verified}.}
    \label{tab:RQ1-result-verified}
    \footnotesize
    \begin{tabular}{llc}
        \toprule
        Method & Backbone & FP Rate \\
        \midrule
        Zero-Shot-Plus & GPT-4/GPT-4o & 14.3\% (\(\uparrow\) 391.61\%) \\
        \textsc{LIBRO} & GPT-4o & 17.8\% (\(\uparrow\) 294.94\%) \\
        OpenHands & Claude 3.5 Sonnet & 27.7\% (\(\uparrow\) 153.79\%) \\
        \textsc{Otter} & GPT-4o & 31.6\% (\(\uparrow\) 122.47\%) \\
        \textsc{Issue2Test} & GPT-4o-mini & 33.33\% (\(\uparrow\) 110.92\%) \\
        \textsc{Otter++} & GPT-4o & 37.4\% (\(\uparrow\) 87.97\%) \\
        \textsc{AssertFlip} & GPT-4o & 45.5\% (\(\uparrow\) 54.51\%) \\
        Amazon Q & Amazon Bedrock & 51.0\% (\(\uparrow\) 37.84\%) \\
        OpenHands & GPT-5-mini & 62.4\% (\(\uparrow\) 12.66\%) \\
        \midrule
        \textbf{\toolname{}} & GPT-5-mini & \textbf{70.30\%} \\
        \bottomrule
    \end{tabular}
\end{table}

\begin{table}[t]
    \centering
    \caption{Comparison of \toolname{} with baselines on \textsc{SWT-bench-lite}.}
    \label{tab:RQ1-result-lite}
    \footnotesize
    \begin{tabular}{llc}
        \toprule
        Method & Backbone & FP Rate \\
        \midrule
        AutoCodeRover & GPT-4 & 9.1\% (\(\uparrow\) 542.09\%) \\
        Zero-Shot-Plus & GPT-4/GPT-4o & 9.4\% (\(\uparrow\) 521.60\%) \\
        \textsc{SWE-Agent} & GPT-4o mini & 9.8\% (\(\uparrow\) 496.22\%) \\
        \textsc{SWE-Agent} & Claude 3.5 Sonnet & 12.3\% (\(\uparrow\) 375.04\%) \\
        \textsc{Aider} & GPT-4 & 12.7\% (\(\uparrow\) 360.08\%) \\
        \textsc{LIBRO} & GPT-4o & 14.1\% (\(\uparrow\) 314.40\%) \\
        \textsc{SWE-Agent} & GPT-4 & 15.9\% (\(\uparrow\) 267.48\%) \\
        \textsc{SWE-Agent+} & GPT-4 & 18.5\% (\(\uparrow\) 215.84\%) \\
        \textsc{Otter} & GPT-4o & 23.33\% (\(\uparrow\) 150.45\%) \\
        OpenHands & Claude 3.5 Sonnet & 28.3\% (\(\uparrow\) 106.47\%) \\
        \textsc{Otter++} & GPT-4o & 29.0\% (\(\uparrow\) 101.48\%) \\
        \textsc{e-Otter} & GPT-4o & 29.0\% (\(\uparrow\) 101.48\%) \\
        \textsc{Otter++} & Claude 3.7 Sonnet & 30.4\% (\(\uparrow\) 92.20\%) \\
        \textsc{Issue2Test} & GPT-4o-mini & 30.43\% (\(\uparrow\) 92.01\%) \\
        \textsc{e-Otter} & Claude 3.7 Sonnet & 36.0\% (\(\uparrow\) 62.31\%) \\
        \textsc{AssertFlip} & GPT-4o & 38.0\% (\(\uparrow\) 53.76\%) \\
        OpenHands & GPT-5-mini & 38.0\% (\(\uparrow\) 53.76\%) \\
        Amazon Q & Amazon Bedrock & 39.9\% (\(\uparrow\) 46.44\%) \\
        \textsc{e-Otter++} & GPT-4o & 40.2\% (\(\uparrow\) 45.35\%) \\
        \midrule
        \textbf{\toolname{}} & GPT-5-mini & \textbf{58.43\%} \\
        \bottomrule
    \end{tabular}
\end{table}

\textbf{Experimental Results.}
Table~\ref{tab:RQ1-result-verified} and \ref{tab:RQ1-result-lite} present the overall comparison results. \toolname{} achieves the best FP Rate on both benchmarks, reaching 58.43\% on \textsc{SWT-bench-lite} and 70.30\% on \textsc{SWT-bench-verified}. On \textsc{SWT-bench-lite}, this result is substantially higher than all compared baselines, where the strongest baseline reaches 38.0\%. On \textsc{SWT-bench-verified}, \toolname{} also outperforms the best baseline result of 62.4\%.

The advantage of \toolname{} is also consistent when compared with recent issue-to-test generation methods. For example, it improves over \textsc{Issue2Test} by 92.01\% on \textsc{SWT-bench-lite} and 110.92\% on \textsc{SWT-bench-verified}, and over \textsc{Otter++} by 119.08\% and 118.32\%, respectively. Compared with \textsc{AssertFlip}, which generates passing tests before inverting assertions, \toolname{} still yields clear improvements on both datasets. Overall, these results indicate that the proposed multi-stage framework is more effective than directly generating tests from retrieved context or relying primarily on post-hoc refinement.

\textbf{Effectiveness across repositories.}
Tables~\ref{tab:RQ1-result-repo-analysis-lite} and~\ref{tab:RQ1-result-repo-analysis-verified} present the results across different projects. 
\toolname{} shows stable performance across a diverse set of projects rather than concentrating its gains on only a small subset of instances. 
On \textsc{SWT-bench-lite}, it achieves reproduction rates above 50\% on several major repositories, including django (64.6\%), sympy (63.4\%), and scikit-learn (68.4\%). 
On \textsc{SWT-bench-verified}, the same trend remains, with particularly strong results on django (74.1\%), sympy (74.0\%), matplotlib (71.9\%), scikit-learn (91.7\%), and astropy (75.0\%).
The results also show that performance is not uniform across repositories.
Some low rates should be interpreted cautiously because they are computed from very few instances.
For example, flask has only one instance in SWT-Bench-Lite, and Issue2Test also gets 0\% on it.

\begin{figure}[!htb]
    \centering
    \includegraphics[width=0.48\textwidth]{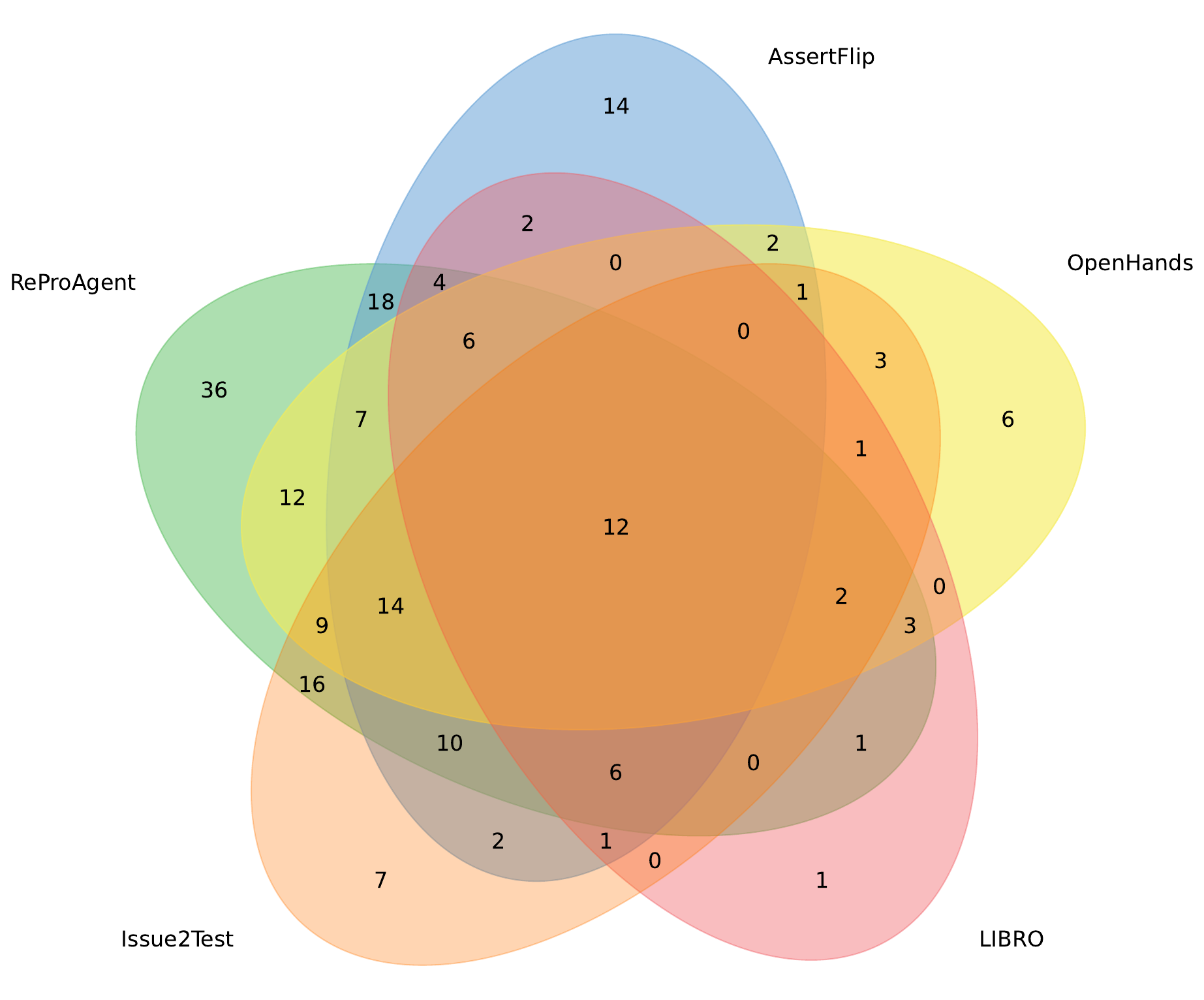}
    \caption{Overlap analysis against baselines}
    \label{fig:venn_digram}
\end{figure}

\textbf{Overlap Analysis.}
To further understand whether the gain of \toolname{} mainly comes from solving the same easy instances more reliably, we analyze the overlap of successful cases between \toolname{} and representative baselines. 
Figure~\ref{fig:venn_digram} presents a overlap analysis among \toolname{}, \textsc{AssertFlip}, OpenHands, \textsc{LIBRO}, and \textsc{Issue2Test} on \textsc{SWT-bench-lite}. The results show that \toolname{} uniquely reproduces 36 issues that are not reproduced by any of the four baselines. 
This indicates that the improvement of \toolname{} is not limited to solving already-easy cases more reliably; instead, it expands the set of issues for which valid reproduction tests can be generated.

\finding{1}{\toolname{} performs best on both \textsc{SWT-bench-lite} and \textsc{SWT-bench-verified}, reaching 58.43\% and 70.30\%, respectively, and it uniquely reproduces 36 issues not reproduced by any of the four representative baselines on \textsc{SWT-bench-lite}.}

\begin{table}[t]
    \centering
    \footnotesize
    \caption{Distribution of results across different projects on SWT-bench-lite}
    \label{tab:RQ1-result-repo-analysis-lite}
    \begin{tabular}{lccc}
        \toprule
        \textbf{Repository} & \textbf{Total Issues} & \textbf{Reproduced} & \textbf{Rate} \\
        \midrule
        django/django                  & 113 &  73 &  64.6\% \\
        sympy/sympy                    &  71 &  45 &  63.4\% \\
        matplotlib/matplotlib          &  23 &  11 &  47.8\% \\
        scikit-learn/scikit-learn      &  19 &  13 &  68.4\% \\
        pytest-dev/pytest              &  11 &   3 &  27.3\% \\
        sphinx-doc/sphinx              &  11 &   2 &  18.2\% \\
        pydata/xarray                  &   5 &   2 &  40.0\% \\
        astropy/astropy                &   4 &   3 &  75.0\% \\
        mwaskom/seaborn                &   4 &   2 &  50.0\% \\
        pylint-dev/pylint              &   3 &   1 &  33.3\% \\
        pallets/flask                  &   2 &   0 &   0.0\% \\
        psf/requests                   &   1 &   1 & 100.0\% \\
        \midrule
        \textbf{Total} & \textbf{267} & \textbf{156} & \textbf{58.43\%} \\
        \bottomrule
    \end{tabular}%
\end{table}

\begin{table}[t]
    \centering
    \footnotesize
    \caption{Distribution of results across different projects on SWT-bench-verified}
    \label{tab:RQ1-result-repo-analysis-verified}
    \begin{tabular}{lccc}
        \toprule
        \textbf{Repository} & \textbf{Total Issues} & \textbf{Reproduced} & \textbf{Rate} \\
        \midrule
        django/django                  & 216 & 160 &  74.1\% \\
        sympy/sympy                    &  73 &  54 &  74.0\% \\
        matplotlib/matplotlib          &  32 &  23 &  71.9\% \\
        sphinx-doc/sphinx              &  28 &   6 &  21.4\% \\
        scikit-learn/scikit-learn      &  24 &  22 &  91.7\% \\
        astropy/astropy                &  16 &  12 &  75.0\% \\
        pydata/xarray                  &  15 &  10 &  66.7\% \\
        pytest-dev/pytest              &  15 &  11 &  73.3\% \\
        pylint-dev/pylint              &   5 &   2 &  40.0\% \\
        psf/requests                   &   4 &   3 &  75.0\% \\
        mwaskom/seaborn                &   2 &   0 &   0.0\% \\
        pallets/flask                  &   1 &   0 &   0.0\% \\
        \midrule
        \textbf{Total} & \textbf{431} & \textbf{303} & \textbf{70.30\%} \\
        \bottomrule
    \end{tabular}%
\end{table}

\subsection{RQ2: Generalizability across Different LLMs}

\textbf{Experimental Design.}
To evaluate whether \toolname{} generalizes across different backbone models, we further implement \toolname{} with three open-source LLMs: DeepSeek-V3.2~\cite{deepseekv3}, GLM-4.6~\cite{glm4_5}, and Qwen3-Coder.
These models differ in scale and training emphasis, while all provide strong support for code generation and agentic reasoning.
We run the same \toolname{} pipeline with each backbone on both \textsc{SWT-bench-lite} and \textsc{SWT-bench-verified}, and use FP Rate as the evaluation metric.

\begin{table}[t]
    \centering
    \caption{Performance across different backbone LLMs}
    \label{tab:RQ2-result}
    \begin{tabular}{lccc}
        \toprule
        \multirow{2}{*}{\textbf{LLM}} 
            & \multicolumn{3}{c}{\textbf{FP Rate}} \\
        \cmidrule(lr){2-4}
            & \textsc{SWT-bench-lite} 
            & \textsc{SWT-bench-verified} 
            & \textbf{Average} \\
        \midrule
        GPT-5-mini
            & \textbf{58.43\%}
            & \textbf{70.30\%}
            & \textbf{64.37\%} \\
        DeepSeek-V3.2
            & 49.64\%
            & 53.58\%
            & 52.11\% \\
        GLM-4.6
            & 48.19\%
            & 52.66\%
            & 50.43\% \\
        Qwen3-Coder
            & 56.88\%
            & 60.05\%
            & 58.47\% \\
        \midrule
        \textbf{Average}
            & \textbf{53.29\%}
            & \textbf{59.15\%}
            & \textbf{56.35\%} \\
        \bottomrule
    \end{tabular}%
\end{table}

\textbf{Experimental Results.}
Table~\ref{tab:RQ2-result} reports the results across different LLM backbones. Overall, \toolname{} achieves strong performance with all four models, obtaining average FP Rates of 64.37\%, 58.47\%, 52.11\%, and 50.43\% with GPT-5-mini, Qwen3-Coder, DeepSeek-V3.2, and GLM-4.6, respectively. Among them, GPT-5-mini achieves the best performance, reaching 58.43\% on \textsc{SWT-bench-lite} and 70.30\% on \textsc{SWT-bench-verified}. Qwen3-Coder ranks second with 56.88\% and 60.05\%, while DeepSeek-V3.2 and GLM-4.6 also obtain competitive results on the two benchmarks.

These results indicate that the effectiveness of \toolname{} does not depend on a single backbone model. Instead, the proposed framework remains effective across LLMs with different model sizes and training characteristics. In particular, even the weaker backbones still achieve competitive FP Rates, suggesting that the gains of \toolname{} mainly come from the framework design rather than from a specific model alone.

The stronger performance of GPT-5-mini is likely due to its stronger coding, long-context reasoning, and tool-use capabilities, which better support repository understanding, structured reasoning, and execution-guided test generation. Nevertheless, the overall trend shows that \toolname{} transfers well across different advanced LLMs.

\finding{2}{\toolname{} generalizes well across different backbone LLMs, achieving average FP Rates of 64.37\%, 58.47\%, 52.11\%, and 50.43\% with GPT-5-mini, Qwen3-Coder, DeepSeek-V3.2, and GLM-4.6, respectively. Across the four backbones, the overall average FP Rate reaches 56.35\%.}

\subsection{RQ3: Contribution of Different Components}

\textbf{Experimental Design.}
To examine the contribution of each stage in \toolname{}, we conduct an ablation study by removing each stage in turn from the full framework. Since full ablation with GPT-5-mini would require substantially higher inference cost, we perform this study with Qwen3-Coder, the strongest open-source backbone in RQ2. Specifically, we evaluate four core stages, including hierarchical bug localization, execution path-based root cause analysis, test generation planning, and feedback iteration with triadic review. The resulting variants are evaluated on both \textsc{SWT-bench-lite} and \textsc{SWT-bench-verified} using FP Rate.

\begin{table}[t]
    \centering
    \caption{Impact of different stages in \toolname{} with Qwen3-Coder}
    \label{tab:RQ3-result}
    \begin{tabular}{lcc}
        \toprule
        \textbf{Variant} 
            & \textsc{SWT-bench-lite} 
            & \textsc{SWT-bench-verified} \\
        \midrule
        w/o Hierarchical Bug Localization 
            & 47.83\% 
            & 55.89\% \\
        w/o Execution Path-based Root Cause Analysis 
            & 53.26\% 
            & 56.58\% \\
        w/o Test Generation Planning 
            & 52.17\% 
            & 57.74\% \\
        w/o Feedback Iteration and triadic review 
            & 18.84\% 
            & 22.86\% \\
        \midrule
        \textbf{Ours (Full Method, Qwen3-Coder)} 
            & \textbf{56.88\%} 
            & \textbf{60.05\%} \\
        \bottomrule
    \end{tabular}%
\end{table}

\textbf{Experimental Results.}
Table~\ref{tab:RQ3-result} presents the ablation results.
Removing any stage leads to a decline in FP Rate on both benchmarks, indicating that all components contribute to the effectiveness of \toolname{}. 
Among them, feedback iteration with triadic review has the largest impact: removing this stage causes FP Rate to drop sharply from 56.88\% to 18.84\% on \textsc{SWT-bench-lite} and from 60.05\% to 22.86\% on \textsc{SWT-bench-verified}. This result highlights the importance of execution-based refinement and review in correcting invalid tests and ensuring that generated tests faithfully capture the intended fail-to-pass behavior.
The other three stages also provide consistent benefits. Removing hierarchical bug localization reduces FP Rate to 47.83\% on \textsc{SWT-bench-lite} and 55.89\% on \textsc{SWT-bench-verified}, suggesting that fine-grained localization helps identify code regions that are more relevant to the reported issue. Removing execution path-based root cause analysis lowers FP Rate to 53.26\% and 56.58\%, respectively, showing that structured reasoning over fault propagation improves downstream test construction. Removing test generation planning further decreases FP Rate to 52.17\% and 57.74\%, indicating that explicit planning provides useful guidance even after the preceding analysis stages.

As bug localization plays an important role in providing suspicious code for subsequent stages, we further analyze how its correctness affects the final fail-to-pass performance.
Table~\ref{tab:localization-sensitivity} groups instances according to whether \toolname{} correctly localizes the buggy code. 
The results show that correct localization substantially improves the final FP rate, increasing it from 50.7\% to 67.8\% on SWT-Bench-Lite and from 60.8\% to 85.3\% on SWT-Bench-Verified. 
This confirms that localization quality is an important factor in generating effective reproduction tests. 
Meanwhile, incorrectly localized cases still achieve non-trivial FP rates, suggesting that later stages can partially recover through additional context retrieval and execution-path reasoning. 
These results also indicate that \toolname{} can further benefit from stronger localization modules in future work.

\begin{table}[t]
\centering
\caption{Sensitivity of \toolname{} to localization correctness.}
\label{tab:localization-sensitivity}
\begin{tabular}{lcc}
\toprule
Dataset & Localization & FP Rate \\
\midrule
SWT-Bench-Lite & Correct & 67.8\% (82/121) \\
SWT-Bench-Lite & Incorrect & 50.7\% (74/146) \\
SWT-Bench-Verified & Correct & 85.3\% (145/170) \\
SWT-Bench-Verified & Incorrect & 60.8\% (158/260) \\
\bottomrule
\end{tabular}
\end{table}

\finding{3}{All four stages contribute positively to \toolname{}, e.g.,  with feedback iteration with triadic review having the largest effect. Removing this stage causes the FP Rate to drop from 56.88\% to 18.84\% on \textsc{SWT-bench-lite} and from 60.05\% to 22.86\% on \textsc{SWT-bench-verified}.
}

\section{Discussion}
\subsection{Reproduction Tests in Issue Repair}

Issue reproduction tests are valuable for automated repair because they serve as a critical criterion for assessing whether a generated patch is actually correct and supporting feedback-driven patch refinement.
Motivated by this role, we further investigate how issue reproduction tests generated by \toolname{} can support downstream repair from two perspectives: improving existing repair frameworks and enabling feedback-driven iterative repair.

\textbf{Benefit to Existing Repair Frameworks}.
We integrate \toolname{} into two representative repair frameworks: agent-based \textsc{SWE-agent}~\cite{sweagent}, and workflow-based \textsc{Agentless}~\cite{agentless}. 
For \textsc{SWE-agent}, we inject tests from \toolname{} into its prompts.
For \textsc{Agentless}, we use tests from \toolname{} in the patch filtering stage alongside its original regression tests.
Considering the cost of running full repair pipelines, we conduct this analysis with Qwen3-Coder, the open-source backbone in RQ2.

\begin{table}[t]
    \centering
    \caption{Effect of reproduction tests on existing repair work}
    \label{tab:discussion-repair-framework}
    \begin{tabular}{lccc}
        \toprule
        \multirow{2}{*}{\textbf{Framework}} 
            & \multirow{2}{*}{\textbf{Dataset}}
            & \multicolumn{2}{c}{\textbf{Patch Pass Rate}} \\
        \cmidrule(lr){3-4}
            & & Before & After \\
        \midrule
        \multirow{2}{*}{\textsc{SWE-agent} + \texttt{Qwen3-Coder}}
            & \textsc{SWE-bench-lite}
                & 143 / 300
                & 153 / 300 \\
            & \textsc{SWE-bench-verified}
                & 317 / 500
                & 327 / 500 \\
        \midrule
        \multirow{2}{*}{\textsc{Agentless} + \texttt{Qwen3-Coder}}
            & \textsc{SWE-bench-lite}
                & 93 / 300
                & 100 / 300 \\
            & \textsc{SWE-bench-verified}
                & 188 / 500
                & 201 / 500 \\
        \bottomrule
    \end{tabular}%
\end{table}

As shown in Table~\ref{tab:discussion-repair-framework}, the generated reproduction tests consistently improve repair performance across both frameworks.
On \textsc{SWE-bench-lite}, the number of successfully resolved issues increases from 143 to 153 for \textsc{SWE-agent} and from 93 to 100 for \textsc{Agentless}.
On \textsc{SWE-bench-verified}, the corresponding numbers rise from 317 to 327 and from 188 to 201, respectively. 
These improvements suggest that reproduction tests provide useful signals, helping better align generated patches with issue semantics and offering stronger behavioral criteria for patch selection.

\textbf{Feasibility of End-to-End Repair with Reproduction Tests}.
We further investigate whether reproduction tests can be used as feedback signals in an end-to-end repair loop. 
To this end, we construct a simple iterative repair pipeline in which the model first identifies suspicious files, then generates candidate patches, verifies them against the generated reproduction tests, and performs another repair iteration if verification fails, up to 5 iterations.
\begin{table}[t]
    \centering
    \caption{Effect of reproduction tests on iterative repair}
    \label{tab:discussion-repair-feedback}
    \begin{tabular}{lccc}
        \toprule
        \multirow{2}{*}{\textbf{Method}} 
            & \multicolumn{3}{c}{\textbf{Patch Pass Rate}} \\
        \cmidrule(lr){2-4}
            & \textsc{SWE-bench-lite}
            & \textsc{SWE-bench-verified}
            & \textbf{Total} \\
        \midrule
        Localization + Repair
            & 99/300 (33.0\%)
            & 194/500 (38.8\%)
            & 293 \\
        \midrule
        \textbf{Localization + Repair + Feedback}
            & \textbf{120/300 (40.0\%)}
            & \textbf{226/500 (45.2\%)}
            & \textbf{346} \\
        \bottomrule
    \end{tabular}%
\end{table}

Table~\ref{tab:discussion-repair-feedback} shows that reproduction-test feedback consistently increases the number of successful repair outcomes on both benchmarks. On \textsc{SWE-bench-lite}, the number of successful cases increases from 99 to 120. On \textsc{SWE-bench-verified}, the corresponding number rises from 194 to 226. In total, the number of successful cases across the two benchmarks increases from 293 to 346. 
These results indicate that reproduction tests can serve not only as final validators, but also as actionable execution-time feedback for repair. 
Although the current pipeline is intentionally simple, the observed gains highlight the practical promise of coupling repair with reproduction-test feedback.

\summary{Issue reproduction tests generated by \toolname{} consistently improve the effectiveness of existing repair frameworks and also serve as actionable feedback in iterative repair, increasing the number of successful cases from 99 to 120 on \textsc{SWE-bench-lite} and from 194 to 226 on \textsc{SWE-bench-verified}.}

\subsection{Cost Analysis}

Since \toolname{} is an agentic framework with iterative tool use, we further analyze its inference cost under the primary GPT-5-mini setting.
For comparison, we collect the reported average per-instance costs of representative baselines if such information is available in the original papers.
As shown in Table~\ref{tab:cost-analysis}, \toolname{} costs \$0.14 per instance on average.
This is slightly higher than OpenHands with GPT-5-mini, but lower than several GPT-4o- or Claude-based reproduction-test generation baselines.
Because different methods use different backbone models and pricing schemes, we do not use monetary cost as a direct superiority claim.
Instead, the result shows that \toolname{} remains affordable in practice while providing substantially higher FP Rate under the same GPT-5-mini backbone.
We also examine the number of feedback iterations used by successful cases.
Among all successfully reproduced instances, the average number of feedback iterations is 1.29, and 90\% of successful cases are resolved within five iterations.
This observation suggests that future work can explore adaptive iteration budgeting or early-stopping strategies to further reduce cost while preserving reproduction effectiveness.

\begin{table}[t]
    \centering
    \caption{Average cost per instance}
    \label{tab:cost-analysis}
    \begin{tabular}{llc}
        \toprule
        \textbf{Method} & \textbf{Backbone} & \textbf{Cost} \\
        \midrule
        \textsc{Otter++} & GPT-4o & \$1.80 \\
        \textsc{Otter++} & Claude-3.7-Sonnet & \$2.75 \\
        \textsc{AssertFlip} & GPT-4o & \$1.00 \\
        \textsc{Issue2Test} & Claude-3.5-Sonnet & \$0.66 \\
        OpenHands & GPT-5-mini & \$0.11 \\
        \toolname{} & GPT-5-mini & \$0.14 \\
        \bottomrule
    \end{tabular}%
\end{table}

\summary{\toolname{} remains affordable in practice, costing \$0.14 per instance on average. Most successful cases require only a small number of feedback iterations, with an average of 1.29 iterations and 90\% completed within five iterations.}

\subsection{Threats to Validity}

\textbf{Internal Validity.} 
Internal validity concerns potential biases that may affect evaluation fairness and consistency. 
First, LLM inference is stochastic, so repeated runs may yield different results. 
Second, backbone LLMs may respond differently to the same prompts due to differences in reasoning and coding ability. 
To mitigate these threats, we use standardized Docker environments, unified prompts, consistent tool settings, and evaluate \toolname{} across multiple backbone LLMs under the same framework.

\textbf{External Validity.} 
External validity concerns whether our findings generalize beyond the current setting. 
First, our experiments are limited to Python repositories in the \textsc{SWT-bench} datasets, so the results may not generalize to other programming languages. 
However, \toolname{}'s high-level workflow is language-agnostic, and its graph-construction component can be extended using existing tools (e.g., CodeGraphContext~\cite{codegraphcontext} supporting 19 languages).
Second, benchmark issues may not fully represent the complexity and diversity of real-world industrial projects. 
To reduce these threats, we evaluate \toolname{} on two widely used benchmark subsets covering multiple repositories and LLMs.

\section{Conclusion}
\label{sec:conclusion}

This paper presents \toolname{}, a multi-stage agent framework for issue reproduction test generation. By integrating task decomposition and reflection, hybrid code retrieval, and runtime interaction, \toolname{} organizes the generation process into four stages: bug localization, root cause analysis, test planning, and test generation. Experiments on \textsc{SWT-bench-lite} and \textsc{SWT-bench-verified} show that \toolname{} consistently outperforms existing baselines, achieving reproduction rates of 58.43\% and 70.30\%, respectively. \toolname{} also generalizes across multiple LLM backbones and improves downstream issue resolution. For example, when integrated with SWE-agent on \textsc{SWT-bench-lite}, it increases the number of successfully resolved issues from 143 to 153. 
These results demonstrate the effectiveness of structured multi-stage agent design for repository-level issue reproduction test generation.

\bibliographystyle{ACM-Reference-Format}
\bibliography{reference}

\end{document}